\documentclass{ifacconf}

\usepackage{graphicx}      % include this line if your document contains figures
\usepackage{natbib}        % required for bibliography

\usepackage{url}
\usepackage{amsmath,amssymb,amsfonts}

\usepackage[dvipsnames]{xcolor}
\usepackage{comment}
\usepackage{booktabs}   % for \toprule etc.
\usepackage{tabularx}   % flexible-width X columns
\usepackage{array}      % \centering\arraybackslash helpers

\usepackage[caption=false,font=normalsize,labelfont=sf,textfont=sf]{subfig}

\usepackage[nolist,nohyperlinks]{acronym}

\usepackage{siunitx}
\newcommand{\Ns}{N_{\text{s}}}   % number of spacecraft
\newcommand{\Ne}{N_{\text{e}}}   % elements per spacecraft

\usepackage [autostyle, english = british]{csquotes}
\MakeOuterQuote{"}

\usepackage{makecell}

\begin{document}
\begin{acronym}
  \acro{LEO}{Low Earth Orbit}
  \acro{DRA}{Direct Radiating Array}
  \acro{GNC}{Guidance, Navigation, and Control}
  \acro{KPI}{Key Performance Indicator}
  \acro{DSS}{Distributed Satellite System}
  \acro{SIGINT}{Signals Intelligence}
  \acro{LSA}{Logarithmic Spiral Array}
  \acro{ELSA}{Enhanced Logarithmic Spiral Array}
  \acro{MRT}{Maximum Ratio Transmission}
  \acro{HPBW}{Half-Power Beamwidth}
  \acro{LVLH}{Local Vertical Local Horizontal}
  \acro{RAAN}{Right Ascension of the Ascending Node}
  \acro{SLL}{Sidelobe Level}
  \acro{SNR}{Signal-to-Noise Ratio}
  \acro{AOCS}{Attitude and Orbit Control System}
  \acro{D2C}{Direct-to-Cell}
  \acro{SDR}{Software-Defined Radio}
\end{acronym}

\begin{frontmatter}
\vspace{-4em}
© 2025 the authors. This work has been accepted to IFAC for publication under a Creative Commons Licence CC-BY-NC-ND.

\title{Control Requirements for Robust Beamforming in Multi-Satellite Systems\thanksref{footnoteinfo}} 
%Attitude and Position Control Requirements for Coherent Multi-Satellite Beam Patterns
%Control Precision Requirements for Robust Beamforming in Multi-Satellite Arrays

\thanks[footnoteinfo]{This research paper was funded by dtec.bw - Digitalization and Technology Research Center of the Bundeswehr. dtec.bw is funded by the European Union - NextGenerationEU.}

\author[First]{Diego Tuzi} 
\author[First]{Thomas Delamotte} 
\author[First]{Andreas Knopp}

\address[First]{Chair of Signal Processing, University of the Bundeswehr Munich,  85579 Neubiberg Germany (e-mail: paper.sp@unibw.de, \{name.surname\}@unibw.de).}

%\address[Second]{Colorado State University, 
%   Fort Collins, CO 80523 USA (e-mail: author@lamar. colostate.edu)}
%\address[Third]{Electrical Engineering Department, 
%   Seoul National University, Seoul, Korea, (e-mail: author@snu.ac.kr)}

\begin{abstract}                % Abstract of 50--100 words
This work investigates the impact of position and attitude perturbations on the beamforming performance of multi-satellite systems. The system under analysis is a formation of small satellites equipped with direct radiating arrays that synthesise a large virtual antenna aperture. The results show that performance is highly sensitive to the considered perturbations. However, by incorporating position and attitude information into the beamforming process, nominal performance can be effectively restored. These findings support the development of control-aware beamforming strategies that tightly integrate the attitude and orbit control system with signal processing to enable robust beamforming and autonomous coordination.
\end{abstract}

%Satellite control applications, Cooperative systems, Networked systems, Distributed systems, Antenna arrays


\end{frontmatter}
%===============================================================================

\section{Introduction}

The New Space era represents the most significant technological advancement in the space sector in recent decades, enabled by cost reductions in the launch sector, satellite miniaturisation, and the increasing exploitation of \ac{LEO}. This paradigm shift has broadened access to space, enabling industrial and academic entities to develop and validate novel technologies through in-orbit demonstrations~(\cite{kinzel_seamless_2022}).

Within this context, multi-satellite systems are gaining prominence for next-generation missions. When implemented using small satellites, they offer increased fault tolerance, resilience, and system-level reliability due to their distributed architecture. The coordinated operation of multiple standardised and miniaturised satellites can improve the performance of traditional monolithic platforms, while also enabling opportunities for cost-effective production and incremental deployment. Furthermore, this architecture facilitates more efficient utilisation of launch vehicle volume, offering a promising alternative to large deployable antennas, which often involve complex mechanisms and strict volume constraints.

However, these advantages come with increased system-level complexity. Effective operation of multi-satellite systems requires coordinated control of satellite positions and attitudes, commonly referred to as formation flying. This coordination relies on networked control strategies, in which satellites exchange information via inter-satellite links to achieve autonomous, self-organising behaviour in orbit. The \ac{GNC} subsystem, and specifically the \ac{AOCS}, plays a central role in this process by providing attitude and orbit determination using onboard sensors and executing control algorithms to maintain the desired formation layout~(\cite{schilling_small_2021}).

This work focuses on a specific class of multi-satellite systems in which each satellite is equipped with a \ac{DRA}, operating coherently to synthesise a virtual antenna aperture. In these configurations, the satellites operate jointly, enabling enhanced design flexibility. For instance, the \ac{HPBW} of the composite beam can be reduced by increasing inter-satellite spacing, without modifying the onboard antenna characteristics. This improves spatial multiplexing in communication systems and enhances resolution in sensing applications.

A critical requirement for coherent beamforming is synchronisation in time, frequency, and phase. Among these, phase synchronisation is particularly important to achieve constructive interference when multiple transceivers operate simultaneously~(\cite{nanzer_distributed_2021}). In this context, phase synchronisation includes not only classical RF phase coherence, but also geometric phase alignment, namely, compensation for the phase errors introduced by position and attitude perturbations in the satellite formation layout.

The formation layout plays a pivotal role in beam synthesis performance, particularly in mitigating grating lobes when the inter-satellite spacing exceeds the conventional half-wavelength limit. Aperiodic layouts such as the \ac{LSA} and \ac{ELSA}~(\cite{tuzi_satellite_2023,tuzi_distributed_2023}) have been shown to mitigate grating lobes effectively under nominal conditions. However, the performance of such configurations must be evaluated under the position and attitude perturbations typical of \ac{LEO}.

This paper investigates the beamforming performance of multi-satellite systems under position and attitude perturbations, focusing on the impact of geometric phase errors. It is shown that geometric phase alignment, achieved by calibrating the precoding process using position and attitude information from the onboard control sensors, can effectively restore the beam pattern. In applications such as passive remote sensing or \ac{SIGINT}, this information may suffice for offline compensation. In real-time applications such as satellite communications, this information must be used in a closed-loop fashion to update the beamforming process.

Simulation results demonstrate that when geometric phase alignment is applied, the beamforming performance of the \ac{LSA} formation layout is preserved, even in the presence of significant position and attitude perturbations. This implies that accurate position and attitude information can reduce the frequency of corrective manoeuvres, which, from a beamforming perspective, are only necessary when the alignment process fails to maintain predefined \acp{KPI}.

Overall, this work contributes to a deeper understanding of how onboard control subsystems can enable coherent operations and robust beamforming performance in advanced multi-satellite systems.
\section{System Model}

This section introduces the system-level model used to evaluate the beamforming performance of a multi-satellite system under position and attitude perturbations. The analysis assumes an \ac{LSA} formation layout of $\Ns$ identical satellites, each equipped with a planar \ac{DRA}.
For clarity, the analysis adopts the terminology of the transmit-side (e.g., departure direction, precoding). Due to reciprocity, all results and definitions apply equivalently to receive-side configurations.

%%%%%%%%%%%%%%%%%%%%%%%%%%%%%%%%%%%%%%%%%%%%%%%%%%%%%%%%%%%%%%%%%%%%%%%%%%%%
\subsection{Nominal Layout and Perturbations}
%%%%%%%%%%%%%%%%%%%%%%%%%%%%%%%%%%%%%%%%%%%%%%%%%%%%%%%%%%%%%%%%%%%%%%%%%%%%

The evaluation is performed at a fixed epoch $t_0$, where the satellites are placed on circular micro-orbits with identical inclination but slightly different \acp{RAAN}. Each satellite $n = 0, \dots, \Ns - 1$ is associated with its own local nadir-pointing \ac{LVLH} frame $\mathcal{L}_n$.

\begin{figure}[t]
  \centering
  \includegraphics[width=1\columnwidth]{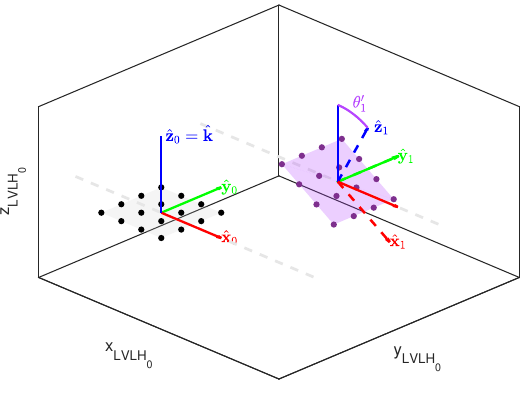}
\caption{Simplified representation of the system model with two satellites and the beam boresight direction. Satellite \(n = 0\) defines the reference frame, while satellite \(n = 1\) is subject to a pitch rotation.}
  \label{fig:model}
\end{figure}

The \ac{LVLH} frame of satellite $n = 0$ is selected as the global reference frame and also serves as the phase reference for the distributed array. 
%This Earth-pointing frame has its $z$-axis directed toward nadir, the $y$-axis opposing the orbit normal, and the $x$-axis completing a right-handed triad aligned with the velocity vector.
A vector expressed in $\mathcal{L}_n$ is mapped to the reference frame $\mathcal{L}_0$ using a translation vector $\mathbf{t}_{n \rightarrow 0} \in \mathbb{R}^{3}$ and a rotation matrix $\mathbf{R}_{n \rightarrow 0} \in \mathrm{SO}(3)$
%\footnote{$\mathrm{SO}(3)=\bigl\{\mathbf C\!\in\!\mathbb R^{3\times3}\,\bigl|\,\mathbf C^{\top}\mathbf C=\mathbf I,\;\det(\mathbf C)=1\bigr\}$, i.e., the set of all $3{\times}3$ orthogonal matrices with unit determinant (proper rotation matrices).}.

Figure~\ref{fig:model} illustrates a simplified system model with two satellites. Satellite \(n = 0\) defines the global \ac{LVLH} frame, while satellite \(n = 1\) is subject to a pitch rotation, with its perturbed element positions shown.

Each satellite hosts an $N_r \times N_c$ planar antenna array lying in its local $xy$-plane. Let $d_x$ and $d_y$ denote the inter-element spacings. The local offset of element $e$ is defined as
\[
\mathbf{d}_e = 
\begin{bmatrix}
(i - \frac{N_r - 1}{2}) d_x \\
(j - \frac{N_c - 1}{2}) d_y \\
0
\end{bmatrix},
\]
with $i = 0, \dots, N_r - 1$, $j = 0, \dots, N_c - 1$, and $e = 0, \dots, \Ne - 1$. The inter-element spacing is conventionally set to $d_x = d_y = \lambda / 2$, with $\lambda$ denoting the signal wavelength.

Nominal satellite positions are generated according to the analytical spiral layout applied in~\cite{tuzi_satellite_2023}, which is designed to suppress grating lobes in the composite radiation pattern.

Deviations from the nominal geometry are modeled by independent, zero-mean truncated normal distributions (\cite{johnson_continuous_1994}) applied to both the translation and rotation of each satellite. The position perturbation is introduced through the random translation vector
\[
\delta\mathbf{t}_n \sim
\mathcal{N}_{\mathrm{trunc}}\left(
\mathbf{0},
\operatorname{diag}(\sigma_x^2, \sigma_y^2, \sigma_z^2),
t_{\max}
\right),
\delta\mathbf{t}_n \in \mathbb{R}^3
\]
which perturbs the satellite center. The parameter $t_{\max}$ ensures that position perturbations do not result in collisions between satellites. Attitude perturbations are modeled by the angle rotation vector
\[
\delta\boldsymbol{\epsilon}_n =
\begin{bmatrix}
\delta\alpha_n \\
\delta\beta_n \\
\delta\gamma_n
\end{bmatrix}
\sim
\mathcal{N}_{\mathrm{trunc}}\left(
\mathbf{0},
\operatorname{diag}(\sigma_\alpha^2, \sigma_\beta^2, \sigma_\gamma^2),
\epsilon_{\max}
\right), 
\delta\boldsymbol{\epsilon}_n \in \mathbb{R}^3
\]
where $\delta\alpha_n$, $\delta\beta_n$, and $\delta\gamma_n$ represent roll, pitch, and yaw perturbations, respectively. The resulting rotation matrix is obtained via the classical $z$–$y$–$x$ Euler angle sequence:
\[
\delta\mathbf{R}_n = R_z(\delta\gamma_n)\, R_y(\delta\beta_n)\, R_x(\delta\alpha_n) \in \mathrm{SO}(3).
\]
The upper bound \(\epsilon_{\max}\) is chosen to exclude full rotations and electromagnetic conditions (e.g. mutual coupling) that are not captured by the beamforming model used in this study.

The perturbed positions of the radiating elements, expressed in the reference frame $\mathcal{L}_0$, are given by
\begin{equation}
\mathbf{p}_{n,e}=
\mathbf{t}'_{n \rightarrow 0} +
\mathbf{R}'_{n \rightarrow 0} \mathbf{d}_e \in \mathbb{R}^3,
\label{eq:positionPert}
\end{equation}
with $\mathbf{t}'_{n \rightarrow 0}=\mathbf{t}_{n \rightarrow 0} + \delta\mathbf{t}_n$ and
$\mathbf{R}'_{n \rightarrow 0} = \mathbf{R}_{n \rightarrow 0} \delta\mathbf{R}_n$.

Nominal positions are recovered by setting $\delta\mathbf{t}_n = \delta\boldsymbol{\epsilon}_n = \mathbf{0}$ for all $n$.

The radiating element positions, expressed in a common reference frame, form the basis for the subsequent power pattern analysis, which is conducted under classical antenna theory assumptions, including narrowband operation and far-field conditions.

%%%%%%%%%%%%%%%%%%%%%%%%%%%%%%%%%%%%%%%%%%%%%%%%%%%%%%%%%%%%%%%%%%%%%%%%%%%%
\subsection{Beamforming and Power Pattern Model}
%%%%%%%%%%%%%%%%%%%%%%%%%%%%%%%%%%%%%%%%%%%%%%%%%%%%%%%%%%%%%%%%%%%%%%%%%%%%

The composite power pattern is derived based on the classical pattern multiplication rule, involving the product of the array geometry, the precoding vector, and the intrinsic radiation pattern of each antenna element (\cite{balanis_antenna_2016}).

The geometry of a single satellite contributes through a per-satellite steering vector. For a departure unit vector $\hat{\mathbf{k}}(\theta,\varphi)$, define $\mathbf{k} = \frac{2\pi}{\lambda} \hat{\mathbf{k}}$, and let
\begin{equation}
\mathbf{a}_n(\theta,\varphi) =
\begin{bmatrix}
e^{j \mathbf{k}^{\top} \mathbf{p}_{n,0}} &
\cdots &
e^{j \mathbf{k}^{\top} \mathbf{p}_{n,\Ne-1}}
\end{bmatrix}^{\top}
\in \mathbb{C}^{\Ne}.
\label{eq:asvPerSat}
\end{equation}
The global multi-satellite steering vector is then given by the vertical concatenation
\[
\mathbf{a}_{\mathrm{MS}}(\theta,\varphi) =
\begin{bmatrix}
\mathbf{a}_0^{\top} & \cdots & \mathbf{a}_{\Ns - 1}^{\top}
\end{bmatrix}^{\top}
\in \mathbb{C}^{\Ns \Ne}.
\]

The precoding vector is based on the \ac{MRT} technique and is computed at the desired direction $(\theta_0,\varphi_0)$. When using nominal or perturbed element positions, the two corresponding \ac{MRT} weights are:
\begin{align}
\mathbf{w}_{\mathrm{nom}} &=
\frac{\left[\mathbf{a}_{\mathrm{MS}}^{\mathrm{nom}}(\theta_0,\varphi_0)\right]^*}
     {\left\lVert \mathbf{a}_{\mathrm{MS}}^{\mathrm{nom}}(\theta_0,\varphi_0) \right\rVert_2},
     \label{eq:precodingNominal}\\
\mathbf{w}_{\mathrm{per}} &=
\frac{\left[\mathbf{a}_{\mathrm{MS}}^{\mathrm{per}}(\theta_0,\varphi_0)\right]^*}
     {\left\lVert \mathbf{a}_{\mathrm{MS}}^{\mathrm{per}}(\theta_0,\varphi_0) \right\rVert_2}, 
     \label{eq:precodingPerturbed}
\end{align}
where the subscript "nom" and "per" refer to nominal and perturbed element positions, respectively, as defined in~\eqref{eq:positionPert}.

Regarding the intrinsic element pattern, the following polar-dependent gain is assumed:
\begin{equation}
f_{\mathrm{el}}(\theta) = \sqrt{G_0} \cos^p \theta, \qquad 0 \le \theta \le \pi/2,
\label{eq:polarPattern}
\end{equation}
where $G_0$ denotes the boresight gain and $p$ is chosen to achieve the desired \ac{HPBW}. 
In the presence of attitude rotation, the element pattern is rotated accordingly. The polar angle between the antenna boresight and the observation direction is given by
\[
\theta_n' = \arccos\left(\hat{\mathbf{z}}_n^{\top} \hat{\mathbf{k}}\right), 
\quad \text{with } \hat{\mathbf{z}}_n = \mathbf{R}'_{n \rightarrow 0} 
\begin{bmatrix} 0 & 0 & 1 \end{bmatrix}^{\top}.
\]
Figure~\ref{fig:model} illustrates this effect for satellite $n = 1$, showing the resulting polar angle $\theta_1'$.

Each element on satellite \(n\) shares the perturbed gain
\[
g_n = f_{\mathrm{el}}\bigl(\theta_n'\bigr).
\label{eq:gainPerSat}
\]
In the absence of attitude rotation, \(\theta_n'\) reduces to the nominal angle \(\theta_n\).

The resulting composite power pattern of the multi-satellite system is
\begin{equation}
P(\theta,\varphi\,|\,\mathbf{w}) =
\left| \mathbf{w}^{\mathrm{H}} \mathbf{a}_{\mathrm{MS}}(\theta,\varphi) \right|^2
\cdot \frac{1}{\Ns} \sum_{n=0}^{\Ns-1} g_n^2,
\label{eq:powerPattern}
\end{equation}
which serves as the foundation for the performance evaluation presented in this study. 

%%%%%%%%%%%%%%%%%%%%%%%%%%%%%%%%%%%%%%%%%%%%%%%%%%%%%%%%%%%%%%%%%%%%%%%%%%%%
\subsection{Key Performance Indicators}
\label{sec:kpi}
%%%%%%%%%%%%%%%%%%%%%%%%%%%%%%%%%%%%%%%%%%%%%%%%%%%%%%%%%%%%%%%%%%%%%%%%%%%%
%\input{figures/figure2.tex}

The \acp{KPI} used for performance evaluation are based on the composite power patterns defined in~\eqref{eq:powerPattern}, computed in three different ways:
\begin{enumerate}
  \item \textbf{Nominal}: The pattern is computed using the analytical layout without perturbations;
\item \textbf{Perturbed}: The steering vector and single-element gain reflect the perturbed positions, but the precoding vector is computed using the nominal geometry, without access to perturbation information (see~\eqref{eq:precodingNominal});
\item \textbf{Calibrated}: The steering vector and single-element gain reflect the perturbed positions, and the control system provides the necessary information, enabling the precoding vector to compensate for the geometric phase errors introduced by the perturbations (see~\eqref{eq:precodingPerturbed}).
\end{enumerate}

All power patterns are evaluated over a $uv$ coordinate grid, which enables visualization of distortions that cannot be captured in a single azimuth or elevation cut.

The following \acp{KPI} are extracted by comparing the $uv$ maps of the nominal configuration with those of the perturbed and calibrated configurations:

\begin{itemize}
  \item \(\text{Pr}(\Delta G_\text{main} < \SI{1}{\decibel})\): Probability that the difference in main lobe gain, in decibels, between the nominal power pattern and the perturbed or calibrated pattern is less than \SI{1}{\decibel}.
  
  \item \(\text{Pr}(\Delta A_\text{HPBW} < \SI{2}{\percent})\): Probability that the absolute change in the area enclosed by the $-3$~\si{\decibel} contour around the main lobe is less than \SI{2}{\percent}.
  
  \item \(\text{Pr}(\Delta G_\text{SLL} < \SI{1}{\decibel})\): Probability that the increase in maximum sidelobe level, relative to the nominal pattern, is less than \SI{1}{\decibel} in the perturbed or calibrated pattern.
  
  %\item \(\text{Pr}(\Delta \theta_\text{main} < \mathrm{HPBW}/10)\): Probability that the angular displacement of the main lobe peak, measured in the $\theta$ direction, is smaller than \SI{10}{\percent} of the nominal \ac{HPBW}.
\end{itemize}

The performance requirements on position and attitude perturbations will be formulated to ensure that each of the above listed probabilities exceeds \SI{90}{\percent}.
\section{Simulation Results}

Simulations were conducted to evaluate the beamforming performance of various multi-satellite configurations, each comprising a total of 2304 radiating elements. The element pattern was defined as in~\eqref{eq:polarPattern}, with a boresight gain of \(G_0 = \SI{5}{\decibel i}\) and the exponent \(p\) chosen to achieve a \ac{HPBW} of \(70^\circ\).
The system is designed to synthesize a beam with a \SI{3}{\kilo\meter} radius on Earth, assuming a \ac{LEO} orbit at \SI{600}{\kilo\meter} altitude and a center frequency of \SI{1}{\giga\hertz}.

Several system architectures can achieve this target, depending on the number of satellites \(N_s\), the number of radiating elements per satellite \(N_e\), and the inter-satellite spacing. The adopted layout follows the \ac{LSA} configuration, as applied in~\cite{tuzi_satellite_2023}, which is known to suppress grating lobes. Table~\ref{tab:spiral-configs} presents one fundamental trade-off among possible multi-satellite configurations.
Configurations with a larger number of satellites \(N_s\) and fewer elements per satellite \(N_e\) improve beamforming isolation, quantified by the difference \(\Delta G_\text{Main-SLL}\) between the main lobe and the highest \ac{SLL}, which is particularly important in satellite communication applications. However, the associated reduction in minimum inter-satellite distance \(d_\text{sat}^\text{min}\) imposes stringent requirements on formation flying, increasing the need for corrective manoeuvres to avoid collisions. This, in turn, impacts fuel consumption and limits mission duration. 
To address this challenge, recent research has explored fuel-free propulsion strategies, such as electromagnetic forces, that can enable large-scale formations of very small satellites (e.g., palm-sized) flying at close spacing~(\cite{shim_feasibility_2025}).

Although the proposed methodology is general, the simulation requirements are derived from a \ac{D2C} use case. This scenario imposes stringent constraints due to the need for active transmission (where high sidelobes may cause harmful interference), tightly packed formations, and real-time processing (to ensure continuous service). These requirements are notably more demanding than those of applications such as passive sensing or satellite-based \ac{SIGINT}, where larger separations and offline processing are acceptable. The common simulation parameters used for both the representative case and the Monte Carlo analysis are summarized in Table~\ref{tab:simParameters}.

\begin{table}[t]
\centering
\caption{Nominal system configurations}
\begin{tabular}{c c c c c c}
\toprule
$N_s$ & $N_c \times N_r$ & $d_\text{sat}^\text{min}$ (\si{\meter}) & $A_\text{virtual}$ (\si{\meter\squared}) & $\Delta G_\text{Main-Sll}$ (\si{\decibel}) \\
\midrule
4   & $24 \times 24$ & \num{10.91} & \num{328.84} & \num{1.43} \\
9   & $16 \times 16$ & \num{7.31} & \num{437.40} & \num{3.91} \\
16  & $12 \times 12$ & \num{6.14} & \num{594.36} & \num{4.92} \\
36  & $8 \times 8$   & \num{4.11} & \num{659.35} & \num{8.85} \\
64  & $6 \times 6$   & \num{3.02} & \num{663.52} & \num{9.84} \\
144 & $4 \times 4$   & \num{2.04} & \num{701.44} & \num{12.44} \\
256 & $3 \times 3$   & \num{1.54} & \num{721.31} & \num{14.71} \\
576 & $2 \times 2$   & \num{1.03} & \num{727.50} & \num{17.44} \\
\bottomrule
\end{tabular}
\label{tab:spiral-configs}
\end{table}
\begin{table}[t]
\centering
\caption{Simulation parameters}
\label{tab:simParameters}
\begin{tabular}{@{}ll@{}}
\toprule
\textbf{Parameter} & \textbf{Value} \\
\midrule
Orbit altitude & \SI{600}{\kilo\meter} \\
Carrier frequency & \SI{1}{\giga\hertz} \\
Wavelength \(\lambda\) & \SI{0.3}{\meter} \\
Element gain \(G_0\) & \SI{5}{\deci\bel i} \\
Element \ac{HPBW} & \SI{70}{\degree} \\
Translation limit \(t_{\max}\) & \makecell{\SI{2.3}{\meter} (\(N_s = 16\))\\\SI{1.2}{\meter} (\(N_s = 64\))} \\
Rotation limit \(\epsilon_{\max}\) & \SI{45}{\degree} \\
Total radiating elements & 2304 \\
Nominal beam radius (Earth) & \SI{3}{\kilo\meter} \\
Array layout & \ac{LSA} \\
\bottomrule
\end{tabular}
\end{table}
\begin{figure}[t]
  \centering
    \subfloat[]{\includegraphics[width=1\columnwidth]{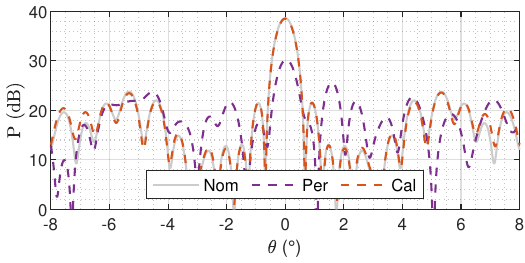}\label{fig:singlerealization_patterncutComparison}}\\
    \subfloat[]{\includegraphics[width=0.49\columnwidth]{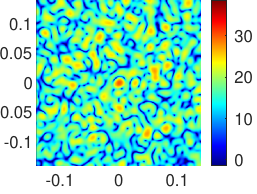}\label{fig:singlerealization_uvPerturbed}}
    \subfloat[]{\includegraphics[width=0.49\columnwidth]{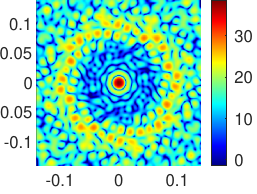}\label{fig:singlerealization_uvCalibrated}}
   \caption{Representative simulation with $N_s = 64$ satellites under translation and rotation perturbations. 
  \protect\subref{fig:singlerealization_patterncutComparison}~$\theta$-cut at $\varphi = 0$ comparing nominal, perturbed, and calibrated patterns.
   \protect\subref{fig:singlerealization_uvPerturbed}~$uv$ power pattern for the perturbed configuration. 
  \protect\subref{fig:singlerealization_uvCalibrated}~$uv$ power pattern after phase-aware calibration.}
    \label{fig:singleRealization}
\end{figure}

\begin{comment}

    \subfloat[]{\includegraphics[width=0.49\textwidth]{figures/singlerealization_uvNominal.pdf}\label{fig:singleRealization_12_uvmapNominal}}%
    \hfill
    \subfloat[]{\includegraphics[width=0.49\textwidth]{figures/singlerealization_uvPerturbed.pdf}\label{fig:singleRealization_13_uvmapPerturbed}}\\[0.4em]
    \subfloat[]{\includegraphics[width=0.8\textwidth]{figures/singlerealization_uvCalibrated.pdf}\label{fig:singleRealization_22_uvmapCalibrated}}
\end{comment}

\subsection{Representative Case}
The representative simulation case considers a multi-satellite system with \(N_s = 64\) satellites and a departure direction defined by \((\theta_0, \varphi_0) = (0^\circ, 0^\circ)\). Position and attitude perturbations are modeled as independent zero-mean random variables with standard deviations \(\sigma_{x} = \sigma_{y} = \sigma_{z} = \lambda/5\) and \(\sigma_{\alpha} = \sigma_{\beta} = \sigma_{\gamma} = \SI{5}{\degree}\).

The simulation results are summarized in Figure~\ref{fig:singleRealization}. The $\theta$-cut at $\varphi = 0$ in Figure~\ref{fig:singleRealization}\subref{fig:singlerealization_patterncutComparison} shows the degradation of the main lobe and the increase in sidelobe levels for the perturbed configuration ("Per") compared to the nominal case ("Nom"). When the perturbations are incorporated into the beamforming process, as in~\eqref{eq:precodingPerturbed}, the calibrated beamforming response ("Cal") closely restores the nominal performance. Consistent observations can be made from the $uv$ domain power patterns. Figure~\ref{fig:singleRealization}\subref{fig:singlerealization_uvPerturbed} shows the distorted pattern under perturbations, while Figure~\ref{fig:singleRealization}\subref{fig:singlerealization_uvCalibrated} displays the result after calibration.

This case highlights the extreme sensitivity of beamforming to position and attitude perturbations. A rotation of only a few degrees, combined with a translation of \(\lambda/5\) (equivalent to \SI{6}{\centi\meter} at \SI{1}{\giga\hertz}), can completely degrade the beam pattern. However, when accurate perturbation information is available, recalculating the beamforming coefficients enables substantial recovery of the nominal performance.

\begin{figure*}[t]
  \centering
  \includegraphics[width=\textwidth]{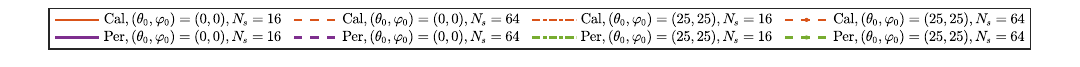}\\[-0.5ex]
  \subfloat[]{\includegraphics[width=.32\textwidth]{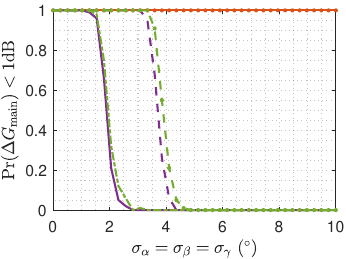}\label{fig:montecarlo_11_deltaG_drot}}\hfill
  \subfloat[]{\includegraphics[width=.32\textwidth]{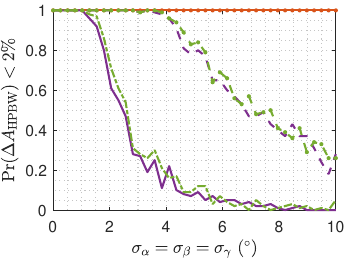}\label{fig:montecarlo_12_deltaAHPBW_drot}}\hfill
  \subfloat[]{\includegraphics[width=.32\textwidth]{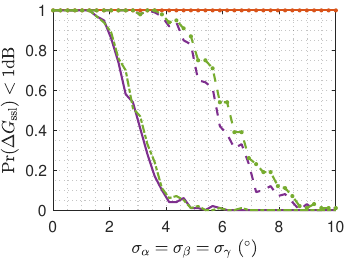}\label{fig:montecarlo_13_deltaGsll_drot}}\\[-0.5ex]

  \subfloat[]{\includegraphics[width=.32\textwidth]{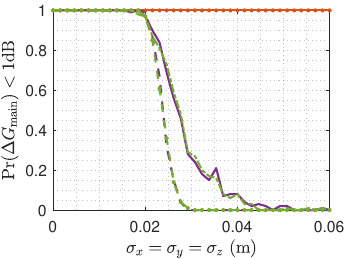}\label      {fig:montecarlo_21_deltaG_dtrans}}\hfill
  \subfloat[]{\includegraphics[width=.32\textwidth]{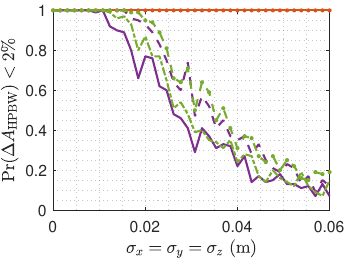}\label{fig:montecarlo_22_deltaAHPBW_dtrans}}\hfill
  \subfloat[]{\includegraphics[width=.32\textwidth]{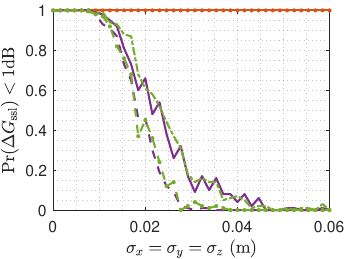}\label{fig:montecarlo_23_deltaGsll_dtrans}}\\[-0.5ex]

  \caption{Monte Carlo simulation results under attitude and position perturbations. 
  The first row (\protect\subref{fig:montecarlo_11_deltaG_drot}–\protect\subref{fig:montecarlo_13_deltaGsll_drot}) shows performance degradation due to attitude perturbations. 
  The second row (\protect\subref{fig:montecarlo_21_deltaG_dtrans}–\protect\subref{fig:montecarlo_23_deltaGsll_dtrans}) shows the impact of position perturbations.}
  \label{fig:mcSimulation}
\end{figure*}

\subsection{Monte Carlo Simulations}

To consolidate and expand the representative case, this section presents a Monte Carlo simulation campaign with \num{100} realizations for each value. The evaluation considers two multi-satellite configurations ($N_s = 16$ and $N_s = 64$) and two beam steering directions: $(\theta_0, \varphi_0) = (0^\circ, 0^\circ)$ and $(\theta_0, \varphi_0) = (25^\circ, 25^\circ)$. The objective is to assess the sensitivity of beamforming performance to position and attitude perturbations, and to evaluate how knowledge of these perturbations can be exploited to restore performance across different steering directions and formation configurations.

Nominal, perturbed, and calibrated $uv$ patterns are compared under independent random perturbations in translation and rotation. For the translation simulation, the standard deviations $\sigma_x = \sigma_y = \sigma_z$ are swept from \SI{0}{\meter} to $\lambda/5$ in \num{40} evenly spaced steps.
The simulations evaluating the effect of rotational perturbations are conducted with $\sigma_{\alpha} = \sigma_{\beta} = \sigma_{\gamma}$ ranging from \SI{0}{\degree} to \SI{10}{\degree}, sampled over \num{40} evenly spaced points. 

Figure~\ref{fig:mcSimulation} summarizes the results. Each subplot shows eight curves, corresponding to the perturbed and calibrated cases, two values of $N_s$, and two departure angles $(\theta_0, \varphi_0)$. The $y$-axis shows the probability that a specific \ac{KPI} is met, while the $x$-axis shows the standard deviation of the random perturbation.

The first row shows the results for attitude perturbations (rotations), while the second row focuses on position perturbations (translations). The columns correspond to the different \acp{KPI} introduced in Section~\ref{sec:kpi}.
The first column, Figures~\protect\subref{fig:montecarlo_11_deltaG_drot} and~\protect\subref{fig:montecarlo_21_deltaG_dtrans}, reports the probability \(\text{Pr}(\Delta G_\text{main} < \SI{1}{\decibel})\). The second column, Figures~\protect\subref{fig:montecarlo_12_deltaAHPBW_drot} and~\protect\subref{fig:montecarlo_22_deltaAHPBW_dtrans}, shows \(\text{Pr}(\Delta A_\text{HPBW} < \SI{2}{\percent})\). The third column, Figures~\protect\subref{fig:montecarlo_13_deltaGsll_drot} and~\protect\subref{fig:montecarlo_23_deltaGsll_dtrans}, shows \(\text{Pr}(\Delta G_\text{SLL} < \SI{1}{\decibel})\).

The perturbed results demonstrate the extreme sensitivity of beamforming performance to position and attitude perturbations. Achieving the specified \acp{KPI} with a \SI{90}{\percent} probability requires individual position accuracy better than \(\lambda/20\), approximately \SI{1.5}{\centi\meter} at \SI{1}{\giga\hertz}, and individual attitude accuracy within a few degrees. In the simulations, position and attitude perturbations are applied independently. However, in practical scenarios, they must be jointly satisfied, making the requirements even more stringent. By contrast, the calibrated results show that beamforming performance is fully restored when geometric phase alignment is achieved, i.e., when the phase errors introduced by the perturbations are compensated. Extending the simulations to larger perturbation levels did not reveal significant degradation in the calibrated patterns.

While the calibrated simulations assume perfect knowledge of the perturbation information, recent advancements in commercial off-the-shelf \acp{SDR} have demonstrated ranging accuracies on the order of \SI{1}{\centi\meter}~(\cite{gardill_towards_2023}), making the required position precision increasingly feasible in practical systems. Furthermore, the required attitude accuracy is comparable to that typically achieved in optical inter-satellite links, where precise beam pointing is essential to maintain high-data-rate laser communication between satellites~(\cite{petermann_distributed_2025}).

Regarding the impact of the beam steering angle, no strong trends are observed across the \acp{KPI}. However, the system configuration with larger \(N_s\) appears more robust to attitude perturbations, as shown in the first row. This trend is not confirmed for position errors: the configuration with fewer satellites shows slightly better robustness in main lobe gain degradation and \ac{SLL} increase (Figures~\protect\subref{fig:montecarlo_21_deltaG_dtrans} and~\protect\subref{fig:montecarlo_23_deltaGsll_dtrans}), while the opposite is observed for \ac{HPBW} area variation (Figure~\protect\subref{fig:montecarlo_22_deltaAHPBW_dtrans}).

\subsection{Individual Position/Attitude Perturbation Impact}
Although all translation \((x, y, z)\) and rotation \((\alpha, \beta, \gamma)\) components are applied simultaneously within each perturbation group, the individual components contribute differently to the geometric phase error, and thus to the beamforming performance.
Both types of perturbations alter the effective position of each radiating element in space. Under narrowband, far-field conditions, this directly impacts the phase contribution of each element as defined in the steering vector~\eqref{eq:asvPerSat}.

In the case of translation, perturbations along the wave propagation direction (the $z$-axis in this setup) have the most direct and significant impact on phase. Even small displacements along $z$ strongly affect phase coherence. Conversely, perturbations along $x$ and $y$ only produce appreciable effects when the beam is steered away from boresight. 

Rotations can also induce significant phase shifts, particularly when the array is physically large, the beam is steered off-boresight, and the rotation axis affects a broad spatial distribution of elements. Pitch rotations (about the \(y\)-axis) and roll rotations (about the \(x\)-axis) effectively displace elements along the \(z\)-axis, introducing phase shifts comparable to those caused by direct translations. Yaw rotations (about the \(z\)-axis), in contrast, primarily affect the beam direction and have limited impact on the projection of element positions onto the departure direction vector \(\hat{\mathbf{k}}\).
The impact of the individual perturbation components is summarized in Table \ref{tab:perturbationEffects}.
\begin{table}[t]
\centering
\caption{Relative impact of individual perturbations on steering vector phase}
\label{tab:perturbationEffects}
\begin{tabular}{@{}lll@{}}
\toprule
\textbf{Perturbation} & \textbf{Strength} & \textbf{Condition for high impact} \\
\midrule
Translation $\delta z$         & +++  & At all steering angles \\
Translation $\delta x, \delta y$ & +    & At large steering angles \\
Rotation $\delta\alpha$ (pitch) & ++   & Wide array in $x$ \\
Rotation $\delta\beta$ (roll)  & ++   & Tall array in $y$ \\
Rotation $\delta\gamma$ (yaw)  & +    & Mainly affects beam direction \\
\bottomrule
\end{tabular}
\end{table}

\section{Conclusion}
%Summarize your key findings. Include important conclusions that can be drawn and further implications for the field. Discuss benefits or shortcomings of your work and suggest future areas for research.

The simulation results confirm that the beamforming performance of multi-satellite systems is highly sensitive to position and attitude perturbations, particularly translations along the \(z\)-axis and pitch and roll rotations. Independent simulations of position and attitude perturbations show that, to meet the most stringent \acp{KPI} defined in this study, position accuracy on the order of \(\lambda/20\) and attitude accuracy within a few degrees are required. In the simulations, position and attitude perturbations are applied independently, however, in practical scenarios, they must be jointly satisfied, making the requirements even more stringent. 

A key insight is that beamforming performance can be substantially recovered when precise position and attitude information is available from onboard sensors and incorporated into the beamforming process. Extended simulations, beyond those presented in this paper, indicate that the calibrated pattern maintains its robustness even under larger perturbations. This suggests that accurate position and attitude information can reduce the frequency of corrective manoeuvres, which, from a beamforming perspective, are only necessary when the alignment process fails to maintain the predefined \acp{KPI}.

For real-time applications such as satellite communications, where maintaining beam performance within target \acp{KPI} is critical for continuous service, the integration of control information into the beamforming process is essential. While the required position accuracy can already be achieved with commercial off-the-shelf \acp{SDR} at the operating frequency considered in this study, moving to higher frequency bands will impose considerably stricter requirements and demand further technological advances. In addition, extending the analysis from a single-epoch evaluation to time-varying scenarios will be necessary to capture the effects of orbital propagation, where beamforming performance, especially the \ac{HPBW}, is expected to fluctuate with the effective virtual antenna aperture of the multi-satellite system and degrade most severely at orbital crossing points.

Taken together, these considerations outline promising directions for advancing control-aware beamforming, in which the \ac{AOCS} is more tightly integrated with the signal processing chain to ensure robust performance and autonomous coordination in next-generation multi-satellite systems.

%\begin{ack}
%You can recognize individuals who provided assistance with your work but who do not meet the definition of authorship
%\end{ack}

\bibliography{ifacconf}             % bib file to produce the bibliography
                                                     % with bibtex (preferred)

\end{document}